\documentclass{aa}  

\usepackage{graphicx}
\usepackage{txfonts}
\begin{document}

   \title{MUSE spectroscopic observations of the Young Massive Cluster NGC1850\thanks{Based on observations collected at 
ESO telescopes under the observing programme 0102.D-0268(A).}}
\titlerunning{MUSE observations of NGC1850}

   \author{A. Sollima
          \inst{1}
           \and
          V. D'Orazi\inst{2,3}
          \and
          R. Gratton\inst{2}
          \and
          R. Carini\inst{4}
           \and
          E. Carretta\inst{1}
           \and
          A. Bragaglia\inst{1}
           \and
          S. Lucatello\inst{2}
         }

   \institute{INAF Osservatorio di Astrofisica e Scienza dello Spazio (OAS),               
   via Gobetti 93/3, 40129 Bologna, Italy\\
              \email{antonio.sollima@inaf.it}
	\and
	 INAF Osservatorio Astronomico di Padova, vicolo dell'Osservatorio 5, 35122 Padova, Italy\\
        \and
	 School of Physics \& Astronomy, Monash University, Clayton VIC 3800, Australia\\
         \and
          INAF Osservatorio Astronomico di Roma, 
	  via Frascati 33, 00078 Monte Porzio Catone, Italy\\
             }

   \date{Received December 15, 2021; accepted March 16, 2022}

 
  \abstract
   {The double cluster NGC1850 in the Large Magellanic Cloud is the nearest Young Massive Cluster of the Local Group with a mass similar to those of Galactic globular clusters.
   Recent studies have revealed an extended morphology of its Main-Sequence turn-off, which can be interpreted as a spread in either age or internal rotation. 
   In spite of its proximity, an accurate spectroscopic determination of its chemical properties is still missing.}
   {We aim at investigating the general chemistry and the kinematics of this stellar system to test whether possible signs of multiple populations are observable in this cluster.}
   {We analyse spectra obtained with MUSE in adaptive optics mode of 1167 stars in both components of this cluster (NGC1850A and NGC1850B).
   Thanks to this dataset, we are able to measure accurate global metallicities, Ba abundances and radial velocities for a sample of 38 Red Supergiants and a guess of the oxygen abundance in the brightest turn-off stars belonging to NGC1850A.}
   {We find an average metallicity of $\langle [M/H]\rangle=-0.31\pm0.01$, a mean Ba abundance of $\langle [Ba/Fe]\rangle=+0.40\pm0.02$ and a systemic radial velocity of $\langle v_{LOS} \rangle=251.1\pm0.3~km~s^{-1}$.
   The dispersion of the radial velocities suggests a dynamical mass of $log(M/M_{\odot})=4.84\pm 0.1$, while no significant systemic rotation is detected.
   We detect a significant bimodality in O~{\sc i} line strength among the turn-off stars of NGC1850A with $\sim$66\% of stars with $[O/Fe]\sim-0.16$ and the rest with no detectable line.
   The majority of O-weak stars populate preferentially the red side of the Main Sequence-turn off and show H lines in emission, suggesting that they are Be stars rotating close to their critical velocity.
   Among normal MSTO stars, red stars have on average broader line profiles than blue ones, suggesting a correlation between colour and rotational velocity.
   }
   {The mean metallicity of this cluster lies at the metal-rich side of the metallicity distribution of the Large Magellanic Cloud 
   following its age-metallicity relation.
   The Ba and O abundances agree with those measured in the bar of this galaxy. 
   The correlation between line broadening and colour suggests that the observed colour spread among turn-off stars can be due to a wide range in rotational velocity covered by these stars.}


   \maketitle
%

\section{Introduction}
\label{intro_sec}

   \begin{figure*}
   \centering
   \includegraphics{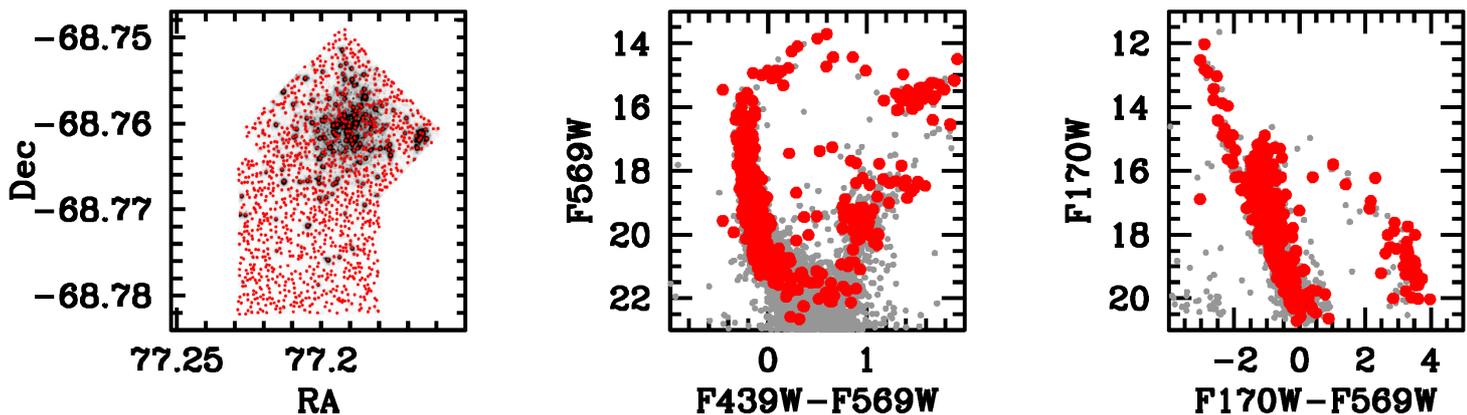}
   \caption{Left panel: map of the analysed MUSE field. Sources with extracted spectra are marked by red dots.
   Middle- and right-panels: (F439W-F569W, F569W) (middle) and (F170W-F569W, F170W) CMDs of NGC1850 (grey points). Target stars are marked by red dots.}
              \label{mapcmd}%
    \end{figure*}

The stellar system NGC1850 is a young \citep[$t_{age}\sim93~Myr$;][]{niederhofer2015} and massive \citep[$M=5.5\times10^{4}~M_{\odot}$;][]{fischer1993} cluster located at the edge of the Large Magellanic Cloud (LMC) bar.
Together with NGC604 in M33 and NGC346 in the Small Magellanic Cloud, it is the 
most massive stellar system of the Local Group younger than 100 Myr \citep{portegieszwart2010}.
An interesting peculiarity of this cluster is that it is projected close to a potential small companion (NGC1850B) with $\sim 1/20$ of the main cluster mass and characterized by a significantly younger age \citep[$t_{age}\sim 4.3~Myr$;][]{gilmozzi1994}.
In the hypothesis that the two clusters are gravitationally bound \citep[as occurring in many other LMC young clusters; ][]{mucciarelli2012,dalessandro2018} this young component would be subject to a strong dynamical friction and will likely merge with the main body of NGC1850 (NGC1850A) in the next $\sim$40 Myr. 

In recent years, new observational evidence has prompted 
a renewed interest in this cluster.
Indeed, deep Hubble Space Telescope (HST) photometric studies have revealed 
multiplicities and spread in the Main Sequence-turn off (MSTO) region of the colour-magnitude diagram (CMD) of a few young ($<1~Gyr$) and intermediate-age ($1-2~Gyrs$) Magellanic Cloud clusters \citep{mackey2008,milone2009}.
Among them, an extended morphology of the MSTO has been found in NGC1850A, with a spread which is not compatible with the typical photometric uncertainties \citep{bastian2016}.
When observed through ultraviolet passbands, a split in the Main Sequence (MS) of this cluster is also apparent \citep{correnti2017}.

Two main interpretations have been proposed to explain the above evidence: 
{\it i)} an age spread of $\sim 30~Myr$ \citep{mackey2008,correnti2017}, similar to that 
predicted in Galactic globular clusters \citep{dercole2008,piotto2015}, or {\it ii)} a 
spread in the rotational velocity of intermediate-mass ($4\div 5~M_{\odot}$) MSTO 
stars \citep{bastian2009}.

All the above evidence implies that, in the hypothesis of a dynamical association between NGC1850 A and B, in the next 10 Gyr this double cluster might evolve into a system with a mass similar to those of present-day globular clusters \citep[$4<log(M/M_{\odot})<6$;][]{baumgardt2018} and 
will end up hosting at least two stellar populations.
It could therefore represents a unique connecting ring between globular clusters 
(where multiple populations are observed several Gyrs after the end of their formation) and young massive clusters (YMCs; where the formation of multiple populations might be still ongoing).

In spite of its importance, direct spectroscopic analyses of individual stars in NGC1850 are extremely rare.
In particular, \citet{fischer1993} analysed echelle spectra of 36 supergiants measuring only their radial velocities.
{Similarly, \citet{kamann2021} derived multi-epoch radial velocities for more than 1000 stars in this cluster to investigate binarity and internal rotation among its members.
Until recent years, all the metallicity estimates of NGC1850 were based on integrated spectroscopy \citep{usher2019} or photometric indices \citep{piatti2019}.
\citet{song2021} determined metallicities for $\sim$100 member stars using medium resolution spectra, but did not analyse the abundance of any specific element.

This observational gap is due to the distance \citep[$d=50~kpc$;][]{niederhofer2015} and intrinsic density of NGC1850 which require an extremely high angular resolution to obtain uncontaminated spectra of individual stars.

In this paper, we analyse more than 1000 individual stars in NGC1850 observed with the Multi Unit Spectroscopic Explorer (MUSE) Integral Field Spectrograph in adaptive optics mode.
This dataset allows to measure the metal content, the Ba abundance and the radial velocities of a subsample of Red Supergiants (RSG) with an excellent accuracy, and to investigate the oxygen abundance and the rotational properties of the brightest MSTO stars.  
The general metal content, the O and Ba abundances are indeed easily measurable even in low-resolution spectra and are crucial in the understanding of the chemical evolution of the gas from which the cluster formed.
These elements are released in the interstellar medium on different timescales by different polluters (SNeII, SNeIa or AGB stars), so 
the relative abundance of these elements provides an indication of the star formation rate and of the role of the various polluters in the chemical enrichment history of the LMC.

\section{Observations and data reduction}
\label{obs_sec}

   \begin{figure*}
   \centering
   \includegraphics{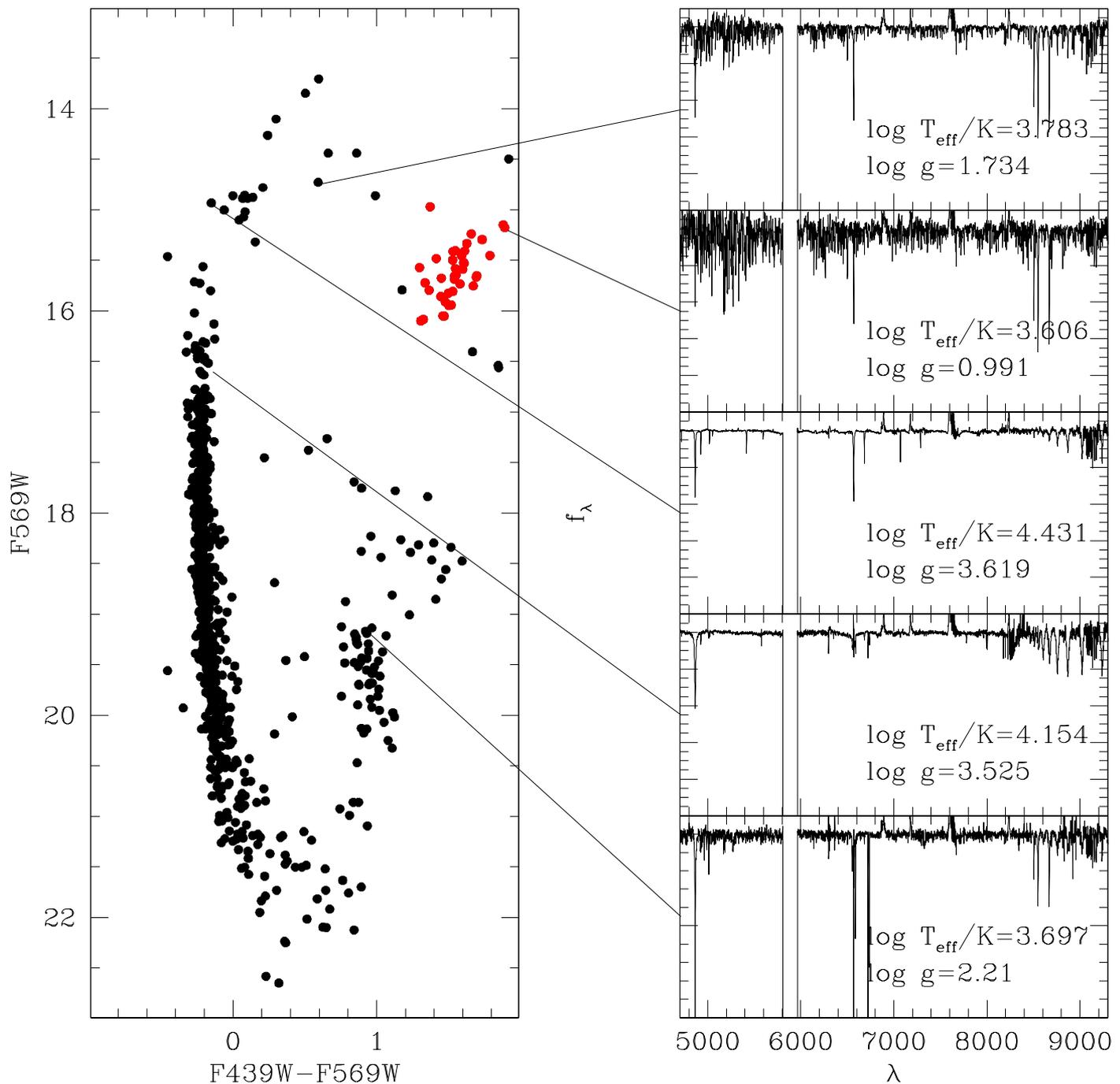}
   \caption{Right panel: from top to bottom, example spectra of a Blue loop star, a RSG, a Blue supergiant, a MS and a Red Giant. 
   Left-panel: (F439W-F569W, F569W) CMD of NGC1850. The positions of the corresponding stars are indicated. The RSGs stars of NGC1850A selected for the kinematic (Sect. \ref{kin_sec}) and chemical (Sect. \ref{met_sec}) analysis are marked by red dots.}
              \label{sho}%
    \end{figure*}

Observations consist of a set of spectral cubes observed with the Integral Field Spectrograph MUSE \citep{bacon2010} at the 
 Very Large Telescope of the European Southern Observatory (Cerro Paranal, Chile) under the observing programme 0102.D-0268(A) (PI: Bastian).
This dataset has been recently used to investigate the relation between binarity and rotation in this cluster \citep{kamann2021}.
The instrument was set up in the wide-field mode using of the ground-layer adaptive optics system.
In this configuration, the instrument field of view is $1\arcmin\times1\arcmin$ in a grid of $288\times 288$ spaxels with a pixel-size of $0.2\arcsec$.
In each spaxel, a medium-resolution grating disperses the light producing a spectrum with a wavelength range between $4700\AA$ and $9350\AA$.
The portion of the spectrum around the sodium doublet between $5805\AA$ and $5965\AA$ has been masked to avoid the strong emission of the laser guide star.
The spectral resolution varies as a function of wavelength ranging from $R\sim 1750$ at $4700\AA$ to $R\sim3750$ at $9300\AA$. 
Observations were performed during six nights between Jan 14 and March 14 2019, with an average seeing of $1.3\arcsec$. 
During each night, $2\times400~s$ exposures were acquired in a field centered on NGC1850A, and $3\times 500~s$ exposures on a partially overlapping field located $\sim 1\arcmin$ away from the cluster centre.
Data cubes were processed and combined using the 2.4 version of the MUSE pipeline \citep{weilbacher2020} yielding a single data cube for each observing block.
At the end of the reduction process, we obtained six data cubes containing in each spaxel a flux- and wavelength-calibrated spectrum.

Spectra were extracted using a specifically developed software. 
As a first step, the images relative to all wavelengths have been stacked together to obtain a high S/N image.
Individual sources have been detected in this image as isolated peaks with intensity above 3$\sigma$ the background noise and no brighter neighbor within 3.4 pixels ($\sim 1~FWHM$).
The sources' centers have been refined as the intensity-weighted x and y positions within 1 FWHM.
A total of 1167 sources have been identified across the entire field of view.
Six bright and isolated sources have been selected and fitted in each wavelength image with Moffat functions with $\beta=2.5$ and variable $\sigma$. 
The average $\sigma$ of these six sources as a function of wavelength has been interpolated using a low-order polynomial, providing a smooth relation between $\lambda$ and $\sigma$.
The corresponding FWHM ranges from 4.2 (at $4700\AA$) to 2.8 pixels (at $9300\AA$).
In the images at each wavelength, the intensity of the detected sources have been then derived by simultaneously fitting all pixels within 1 FWHM from any source with Moffat functions with the appropriate value of $\sigma$ using an iterative PSF-fitting routine.  
For each source, an average background, calculated as the 3$\sigma$-clipped mean of 
pixels contained in an annulus between 4 and 8 FWHM from the considered source and more distant than 1 FWHM from any other source, has been subtracted during the fit. 
In the wavelength ranges encompassing hydrogen lines, this task is complicated by the presence of the intense and spatially variable emission from diffuse gas.
However, the spatial variation of this emission has a typical scale larger than the PSF FWHM, so that 
an effective background subtraction is guaranteed also for stars embedded in gas clouds.

To identify the various targets on a CMD, archive HST images of NGC1850 have been also retrieved and analysed.
They were observed with the Wide Field Planetary Camera 2 (WFPC2) on 1994 March 4 
through the F170W, F439W and F569W filters (prop. ID:5559; PI: Gilmozzi). 
This dataset has been already presented in \citet{gilmozzi1994}.
PSF-fitting photometry has been performed using the \rm{DOLPHOT} software \citep{dolphin2000}.
The WFPC2 pointing has an almost complete overlap with the MUSE field of view, thus providing 
accurate magnitudes for 1131 ($\sim97\%$) of the MUSE targets.
 
The targets cover a wide range of magnitude ($13<F539W<23$) and colour ($-1<F439W-F569W<2$) sampling the population of RSG, blue loop, MS of NGC1850A, the Blue supergiants of NGC1850B and Red Giants of the LMC (see Fig. \ref{mapcmd}).
The S/N ratio of the spectra ranges from 3 (at $F539W\sim22$) to 150 (at $F539W\sim14$).
Examples of the extracted spectra of a few stars are shown in Fig. \ref{sho}

\section{Data analysis}
\label{method_sec}

   \begin{figure}
   \centering
   \includegraphics[width=\hsize]{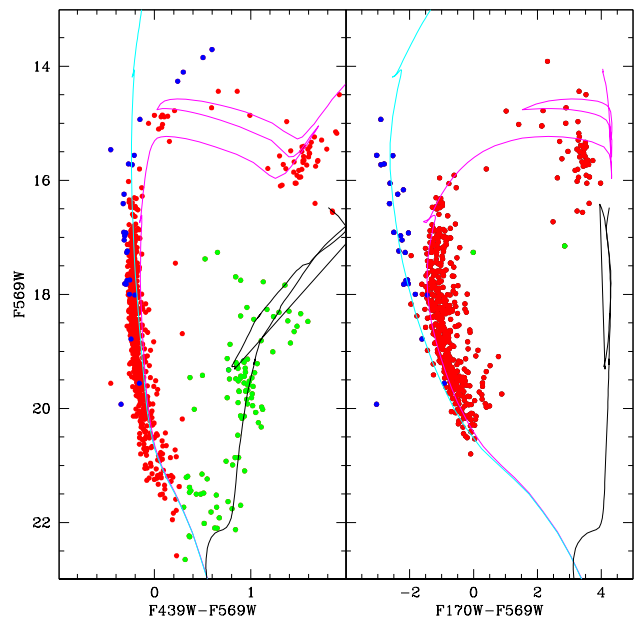}
      \caption{(F439W-F569W, F569W) (left) and (F170W-F569W, F170W) (right) CMDs 
      of NGC1850. Target stars associated to NGC1850A, NGC1850B and the LMC are 
      marked by red, blue and green dots, respectively. The \citet{marigo2008} 
      isochrones with ages of 15 Myr, 90 Myr and 10 Gyr are plotted in both panels with cyan, magenta and black solid lines, respectively.
      We adopted the distance modulus $(m-M)_{0}=18.50$ \citep{niederhofer2015} and the reddening $E(B-V)=0.13$ \citep{gorski2020}.
      }
         \label{isoc}
   \end{figure}

The spectra of target stars have been used to measure their radial velocities and abundances. 
For these purposes, estimates of their effective temperatures and gravities are needed.

These quantities have been calculated through the comparison with suitable theoretical isochrones.
We adopted the set of solar-scaled isochrones of \citet{marigo2008} with a metallicity of Z=0.008 and ages of 15 Myr, 90 Myr and 10 Gyr for NGC1850B, NGC1850A and the LMC, respectively.
Isochrones in the WFPC2 absolute Vegamag photometric system have been transformed into apparent ones using the distance modulus $(m-M)_{0}=18.50$ \citep{niederhofer2015} and reddening E(B-V)=0.13 \citep{gorski2020}.
A reference isochrone has been associated with each star, according to its position in the CMDs: {\it i)} stars within $5\arcsec$ from the center of NGC1850B or F170W-F569W<-2 have been associated with the youngest 15 Myr isochrone, 
{\it ii)} stars with F439W-F569W>0.3 and F569W>17 have been assigned to the oldest 10 Gyr isochrone, and {\it iii)} the remaining stars have been associated to the 90 Myr isochrone (see Fig. \ref{isoc}).
The above criteria unambiguously separate stars belonging to NGC1850A, NGC1850B and the LMC at bright magnitudes ($F569W<18$, including all the stars employed in the chemical and kinematic analysis). 
A certain degree of misassociation is possible at faint magnitudes and blue 
colours ($F569W>18,~F569W-F439W<0.2$). 
However, in this range all isochrones predict very similar colour-temperature and magnitude-gravity relations, so that this effect is not expected to significantly affect the derived stellar parameters\footnote{Since the purpose of the described task is to determine the most appropriate temperature and gravity, 
the adopted criterion is set to associate each star to the most appropriate isochrone, not necessary the one corresponding to the host cluster. 
For instance, the temperature and gravity of a NGC1850A blue straggler star, 
rejuvenated through mass-transfer in binary, are better reproduced by the young isochrone.}.
A distance in the various CMDs between each star and the associated isochrone has been calculated
adopting the metric between colours and magnitudes $(\Delta(F170W-F439W),\Delta(F170W-F569W),\Delta(F439W-F569W),\Delta F170W, \Delta F439W, \Delta F569W)=(0.84,0.88,0.66,1.4,1.19,1.22)$\footnote{Such a metric has been chosen to approximately reproduce the r.m.s. of the various colours and magnitudes in the WFPC2 catalog among stars with all the three available magnitudes.}.   
The temperature and gravity corresponding to the point of the isochrone at the minimum distance have been assigned to the star.

The technique described above requires at least two available magnitudes. 
Unfortunately, some (346) faint targets have one or no WFPC2 magnitudes.
   
To overcome to this problem, a pseudo colour and magnitude have been calculated for all targets from their spectra as 

\begin{eqnarray*}
mag&=&-2.5~log~\sum f_{\lambda}\nonumber\\
col&=&-2.5~log~\frac{\sum\limits_{\lambda<7000\AA} f_{\lambda}}{\sum\limits_{\lambda>7000\AA} f_{\lambda}}\nonumber\\
\end{eqnarray*}

Linear relations linking the various WFPC2 colours and magnitudes with $col$ and $mag$ have been calculated and adopted to derive the missing magnitudes and colours of the faint targets needed to apply the above procedure.

   \begin{figure*}
   \centering
   \includegraphics{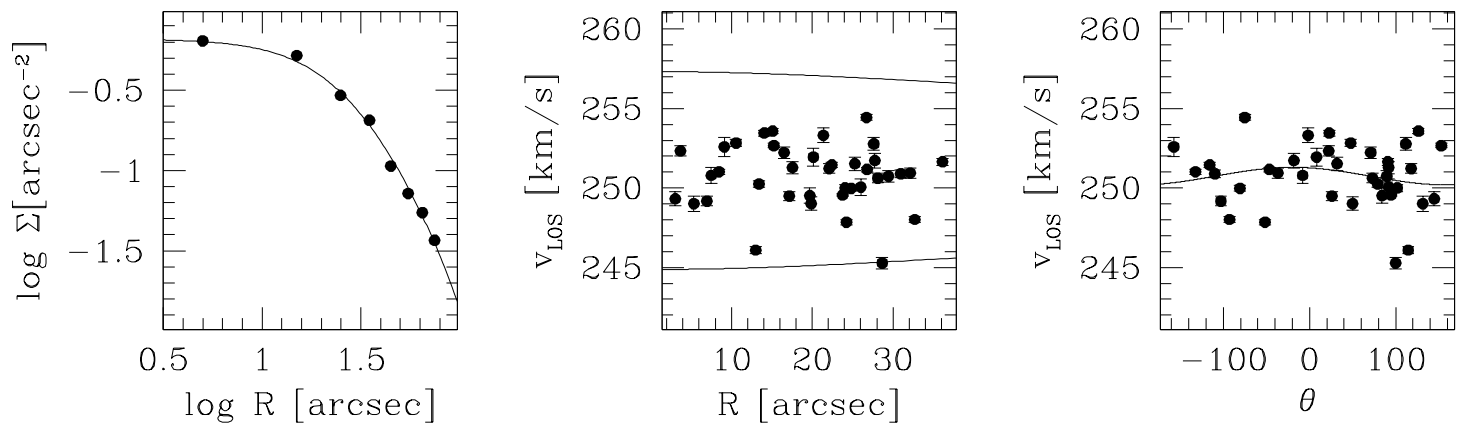}
   \caption{Left panel: projected density distribution of NGC1850A. The best fit \citet{king1966} model is overplotted. 
   Middle-panel: distribution of radial velocities of RSG stars as a function of the distance from the cluster centre. The 3$\sigma$ boundary predicted by the best fit \citet{king1966} model is overplotted.
   Right-panel: distribution of radial velocities of the same sample as a function of the position angle. The best fit sinusoidal curve is overplotted.}
              \label{rot}%
    \end{figure*}

\subsection{Kinematics}
\label{kin_sec}

The radial velocities of target stars have been derived by cross-correlating stellar spectra 
with appropriate synthetic spectra from the Bt-Settl model grid \citep{allard2014}. 
For each target star, a synthetic spectrum with the same temperature and gravity and $[M/H]=-0.5$ has been 
interpolated through the grid and adopted as a template.
Cross-correlation has been performed using the \rm{fxcorr} task within the IRAF\footnote{IRAF is distributed by the National Optical Astronomy
Observatories, which is operated by the Association of Universities for
Research in Astronomy, Inc. (AURA) under cooperative agreement with the National
Science Foundation.} package \citep{tody1986}.
Radial velocities of different epochs have been individually derived and averaged. 
The error on the mean provides an estimate of the associated uncertainty.
Accuracies vary greatly across the CMD: cold ($F439W-F569W>0.8$; corresponding to temperatures $T<5000~K$) stars have 
errors $\sigma_{v}<0.5~km~s^{-1}$ down to $F569W<18$ increasing to 
$\sigma_{v}\sim 2~km~s^{-1}$ at $F569W=20$, while errors as large as 
$\sigma_{v}\sim 5~km~s^{-1}$ can be found among blue ($F170W-F569W<-1.5;~T>20000K$) stars even at bright magnitudes ($F569W<18$).
At fainter magnitudes ($F569W>20$) the S/N drops, leading to errors $\sigma_{v}>10~km~s^{-1}$ regardless of colour.
Given the significant contamination from field LMC stars and the significantly large error in the radial velocities of MS stars, 
we limited the kinematic analysis to a subsample of 38 RSG.
They were selected on the (F439W-F569W, F569W) CMD as those stars with $F439W-F569W>1$ and $15<F569W<16.2$, and are characterized by an extremely accurate radial velocity ($0.1<\sigma_{v}/km~s^{-1}<0.6$).
None of these stars exhibit a significant radial velocity variation among the 6 epochs. 
This means that either they are single stars or that their companion is a low-mass star not significantly affecting the radial systemic velocity.
A star-to-star comparison with the 9 RSGs in common with the sample analysed by \citet{song2021} indicates a difference of $\Delta v_{LOS,this work-S21}=0.06\pm0.30~km/s~(r.m.s.=0.91~km/s)$ and 
$\Delta v_{LOS,this work-S21}=0.17\pm1.04~km/s~(r.m.s.=3.11~km/s)$ including 6 further blue stars in common, indicating no systematics between the two determinations.
The error-weighted average velocity of these stars is $\langle v_{LOS}\rangle=251.1\pm0.3~km~s^{-1}$.
This measure is in agreement with that provided by \citet{fischer1993} ($\langle v_{LOS,F93}\rangle=251.4\pm2.0~km~s^{-1}$) while it is slightly 
higher than that found by \citet{song2021} ($\langle v_{LOS,S21}\rangle=248.9_{-0.5}^{+0.4}~km~s^{-1}$).

The distribution of radial velocities as a function of the position angle has been also used to test the possible presence of systemic rotation (see Fig. \ref{rot}).
Indeed, in the presence of a significant rotation with an axis inclination angle $i\neq 0$, a sinusoidal modulation should be observable.
The best fit with a sine function provides a rotation amplitude of $A=0.5\pm 0.5~km~s^{-1}$.
To test the significance of such a signal, the analysis has been repeated $10^{4}$ times by randomly swapping the 
position angles of stars. Amplitudes as large as the observed one have been obtained in $\sim 52\%$ of extractions indicating that the detected rotation signal is not significant.
This evidence conflicts with the significant rotation signal claimed by \citet{fischer1993} ($A=2.1~km~s^{-1}$ with a 93\% confidence level).

The dynamical mass of NGC1850A has been estimated by fitting the velocity dispersion profile of RSGs with the prediction of an isotropic \citet{king1966} model.
For this purpose, the projected density profile of NGC1850A has been derived by selecting from the entire WFPC2 photometric catalog MS and RSG stars brighter than $F569W<20$ on the (F439W-F569W, F569W) CMD, and calculating the local number density in concentric annuli around the cluster centre.
The density profile has been then fitted with a \citet{king1966} model with central adimensional potential $W_{0}=5$ and core radius $r_{c}=26\arcsec .6$ (corresponding to a half-mass radius of $r_{h}=12.88~pc$ at a distance of 50 kpc; see Fig. \ref{rot}).
The dynamical mass has been then estimated by normalizing the corresponding velocity dispersion profile to the distribution of observed radial velocities, using a maximum-likelihood algorithm \citep[see e.g.][]{pryor1993}.
We derived a dynamical mass of $log(M/M_{\odot})=4.84\pm0.10$, in good agreement with that found by \citet{fischer1993} ($log(M/M_{\odot})_{F93}=4.76\pm0.18$) and \citet{song2021} ($log(M/M_{\odot})_{S21}=4.71_{-0.18}^{+0.04}$).
A similar estimate has been found from photometric studies by \citet{correnti2017} and \citet{mclaughlin2005}($log(M/M_{\odot})=4.86\pm0.10$ in both studies).

\begin{table*}
\caption{Radial velocities, metallicities and Ba abundances of the 38 target RSGs of NGC1850A}             
\label{rsg_tab}      
\centering          
\begin{tabular}{l c c c c c c c c c r}     
\hline\hline       
ID     &  RA        & Dec         & F170W  & F439W  & F569W  & $v_{LOS}$       & log g & $T_{eff}$ & [M/H] & [Ba/Fe]\\
       &            &             &        &        &        & $km~s^{-1}$          &       & $K$ & &\\
\hline                    
RSG 1  & 77.1891513 & -68.7547813 & 18.236 & 16.342 & 14.970 & 249.99$\pm$0.21 & 1.37 & 4820 & -0.19$\pm$0.10 & 0.65$\pm$0.14\\
RSG 2  & 77.1955380 & -68.7609447 &        & 17.031 & 15.149 & 252.33$\pm$0.32 & 0.99 & 4080 & -0.18$\pm$0.08 & 0.55$\pm$0.12\\
RSG 3  & 77.2028529 & -68.7598187 & 18.469 & 17.069 & 15.176 & 253.45$\pm$0.17 & 0.99 & 4106 & -0.24$\pm$0.06 & 0.40$\pm$0.12\\
RSG 4  & 77.1860433 & -68.7579496 & 18.549 & 16.898 & 15.238 & 253.58$\pm$0.16 & 1.12 & 4252 & -0.68$\pm$0.10 & 0.40$\pm$0.11\\
RSG 5  & 77.1762850 & -68.7623850 & 18.572 & 17.029 & 15.294 & 250.72$\pm$0.34 & 1.12 & 4319 & -0.26$\pm$0.05 & 0.35$\pm$0.12\\
RSG 6  & 77.2044573 & -68.7666047 & 18.832 & 16.962 & 15.333 & 247.85$\pm$0.17 & 1.19 & 4414 & -0.30$\pm$0.11 & 0.35$\pm$0.13\\
RSG 7  & 77.1975002 & -68.7758686 & 18.738 & 17.391 & 15.386 & 249.38$\pm$0.13 & 1.05 & 4012 & -0.18$\pm$0.04 & 0.45$\pm$0.13\\
RSG 8  & 77.1986001 & -68.7616196 & 18.887 & 24.077 & 15.390 & 250.79$\pm$0.49 & 1.29 & 4502 & -0.29$\pm$0.09 & 0.50$\pm$0.14\\
RSG 9  & 77.1851371 & -68.7541902 & 18.816 & 16.954 & 15.405 & 252.77$\pm$0.41 & 1.29 & 4605 & -0.32$\pm$0.07 & 0.35$\pm$0.12\\
RSG 10 & 77.1892216 & -68.7534783 & 18.853 & 17.021 & 15.408 & 245.27$\pm$0.36 & 1.19 & 4402 & -0.29$\pm$0.06 & 0.40$\pm$0.13\\
RSG 11 & 77.1915189 & -68.7547468 & 18.941 & 16.945 & 15.413 & 249.55$\pm$0.08 & 1.29 & 4630 & -0.37$\pm$0.07 & 0.40$\pm$0.12\\
RSG 12 & 77.1849157 & -68.7694328 & 18.913 & 17.241 & 15.452 & 250.89$\pm$0.24 & 1.19 & 4209 & -0.25$\pm$0.10 & 0.50$\pm$0.12\\
RSG 13 & 77.1864469 & -68.7623052 & 18.689 & 17.046 & 15.454 & 252.58$\pm$0.62 & 1.27 & 4326 & -0.32$\pm$0.04 & 0.50$\pm$0.14\\
RSG 14 & 77.1706326 & -68.7592967 & 18.773 & 16.896 & 15.481 & 249.25$\pm$0.45 & 1.43 & 4804 & -0.24$\pm$0.07 & 0.45$\pm$0.11\\
RSG 15 & 77.1910025 & -68.7608385 & 18.826 & 17.026 & 15.496 & 249.31$\pm$0.44 & 1.34 & 4240 & -0.34$\pm$0.09 & 0.45$\pm$0.12\\
RSG 16 & 77.1925124 & -68.7512901 & 18.967 & 17.130 & 15.523 & 251.64$\pm$0.22 & 1.27 & 4538 & -0.35$\pm$0.08 & 0.40$\pm$0.14\\
RSG 17 & 77.2093896 & -68.7576391 & 18.959 & 17.140 & 15.530 & 251.52$\pm$0.43 & 1.27 & 4462 & -0.35$\pm$0.06 & 0.50$\pm$0.13\\
RSG 18 & 77.2081946 & -68.7605227 & 19.072 & 17.129 & 15.580 & 251.94$\pm$0.56 & 1.34 & 4438 & -0.31$\pm$0.04 & 0.40$\pm$0.12\\
RSG 19 & 77.1796305 & -68.7628682 & 18.908 & 17.189 & 15.589 & 250.09$\pm$0.23 & 1.34 & 4809 & -0.15$\pm$0.08 & 0.40$\pm$0.14\\
RSG 20 & 77.2002277 & -68.7616722 & 18.386 & 24.046 & 15.622 & 251.58$\pm$0.28 & 1.48 & 4677 & -0.28$\pm$0.07 & 0.30$\pm$0.14\\
RSG 21 & 77.1983491 & -68.7591504 & 19.224 & 17.198 & 15.645 & 252.84$\pm$0.23 & 1.34 & 4444 & -0.33$\pm$0.06 & 0.35$\pm$0.14\\
RSG 22 & 77.1970198 & -68.7569939 & 19.201 & 17.348 & 15.651 & 252.23$\pm$0.37 & 1.27 & 4316 & -0.31$\pm$0.04 & 0.30$\pm$0.13\\
RSG 23 & 77.1981650 & -68.7685052 & 19.386 & 17.197 & 15.654 & 254.44$\pm$0.18 & 1.42 & 4479 & -0.38$\pm$0.07 & 0.50$\pm$0.13\\
RSG 24 & 77.1930351 & -68.7531488 & 19.183 & 17.358 & 15.665 & 250.74$\pm$0.36 & 1.34 & 4332 & -0.34$\pm$0.03 & 0.35$\pm$0.13\\
RSG 25 & 77.1992255 & -68.7538598 & 19.070 & 17.126 & 15.675 & 250.62$\pm$0.30 & 1.42 & 4780 & -0.32$\pm$0.05 & 0.30$\pm$0.12\\
RSG 26 & 77.2026966 & -68.7570936 & 19.103 & 17.232 & 15.689 & 249.02$\pm$0.42 & 1.42 & 4662 & -0.28$\pm$0.09 & 0.35$\pm$0.14\\
RSG 27 & 77.2047532 & -68.7592420 & 19.256 & 17.313 & 15.732 & 249.49$\pm$0.27 & 1.42 & 4558 & -0.30$\pm$0.07 & 0.40$\pm$0.13\\
RSG 28 & 77.2070506 & -68.7667304 & 19.185 & 17.425 & 15.752 & 251.17$\pm$0.14 & 1.34 & 4469 & -0.37$\pm$0.09 & 0.40$\pm$0.13\\
RSG 29 & 77.1925074 & -68.7564521 &        & 17.161 & 15.796 & 251.28$\pm$0.37 & 1.61 & 4619 & -0.33$\pm$0.07 & 0.45$\pm$0.14\\
RSG 30 & 77.1948737 & -68.7576731 & 19.065 & 17.338 & 15.808 & 250.25$\pm$0.21 & 1.51 & 4478 & -0.38$\pm$0.05 & 0.40$\pm$0.12\\
RSG 31 & 77.1825199 & -68.7593800 & 19.000 & 17.323 & 15.824 & 252.66$\pm$0.14 & 1.51 & 4216 & -0.43$\pm$0.10 & 0.40$\pm$0.14\\
RSG 32 & 77.1785855 & -68.7605536 &        & 17.303 & 15.857 & 249.30$\pm$0.35 & 1.51 & 4748 & -0.25$\pm$0.07 & 0.35$\pm$0.13\\
RSG 33 & 77.1922522 & -68.7541042 & 19.436 & 17.391 & 15.912 & 250.06$\pm$0.52 & 1.51 & 4604 & -0.38$\pm$0.09 & 0.30$\pm$0.13\\
RSG 34 & 77.2093007 & -68.7615017 & 19.571 & 17.461 & 15.942 & 253.32$\pm$0.45 & 1.51 & 4576 & -0.29$\pm$0.07 & 0.35$\pm$0.12\\
RSG 35 & 77.1960072 & -68.7681351 & 19.230 & 17.449 & 15.946 & 249.97$\pm$0.18 & 1.51 & 4435 & -0.43$\pm$0.07 & 0.30$\pm$0.14\\
RSG 36 & 77.2131170 & -68.7637558 & 20.026 & 17.511 & 16.050 & 251.72$\pm$0.46 & 1.61 & 4579 & -0.28$\pm$0.09 & 0.35$\pm$0.15\\
RSG 37 & 77.1854110 & -68.7669424 & 19.595 & 17.521 & 16.051 & 251.45$\pm$0.17 & 1.61 & 4546 & -0.35$\pm$0.07 & 0.35$\pm$0.13\\
RSG 38 & 77.1944616 & -68.7558836 &        & 17.409 & 16.084 & 249.51$\pm$0.48 & 1.73 & 4558 & -0.38$\pm$0.07 & 0.40$\pm$0.13\\
\hline      													   
\end{tabular}													   
\end{table*}	    
 
   \begin{figure}
   \centering
   \includegraphics[width=8cm]{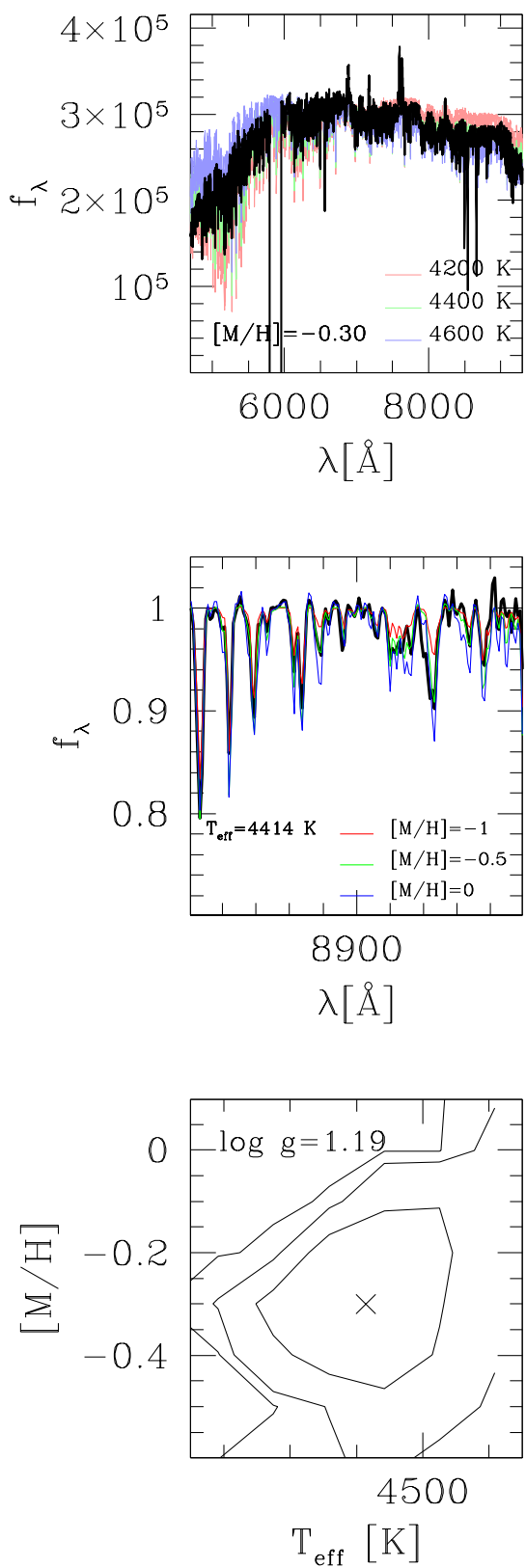}
   \caption{Example of 2D fit in the $T_{eff}-[M/H]$ plane for a sample RSG star. 
   The probability map is shown in the bottom panel. The location of the peak and the 1, 2 and 3$\sigma$ contours are shown.
   The spectrum fit along two projections are shown for different temperatures (top panel) and metallicities (middle panel; in this panel only a small portion of the spectrum is shown for a better readability).}
              \label{mhex}%
    \end{figure}

\subsection{Global metallicity}
\label{met_sec}

The spectra of RSGs in NGC1850A contain a plethora of metal lines across the entire spectrum. Moreover, stellar winds originating from the envelope expansion carry out any eventual angular momentum, thus
preventing any rotation-induced chemical anomalies. They are therefore well suited to estimate the global metallicity of this cluster. 
For this purpose, we adopted a full spectrum fit with synthetic templates.
For each star, the spectra of the six exposures were corrected for Doppler shift and averaged. 
The spectral range ($7000\AA-9000\AA$) has been used in the fit, masking the emission lines, when present.

We adopt the grid of synthetic spectra of Bt-Settl \citep{allard2014}.
This library adopts the solar metallicity of \citet{asplund2009} and a metallicity dependent $\alpha$-enhancement ([$\alpha$/Fe]=+0.4 for $[M/H]\leq-1$, [$\alpha$/Fe]=+0.2 for $-1<[M/H]\leq -0.5$, [$\alpha$/Fe]=0 for $[M/H]>-0.5$).
Synthetic spectra have been degradated at the MUSE resolution and
fluxes have been corrected for dust extinction using the attenuation curve of \citet{cardelli1989}, adopting a reddening E(B-V)=0.13 \citep{gorski2020}. 

In the fit, we kept fixed the gravity to the photometric estimate and searched the minimum $\chi^{2}$ in the $T_{eff}-[M/H]$ plane.
The search has been made using a Powell's conjugate direction algorithm \citep{brent1973} starting from the photometric temperature and a metallicity [M/H]=-0.5 as first guesses.
So, the best fit is obtained with the parameters providing simultaneously a good fit of both the global shape of the continuum and the line depths.
An example of the 2D fit is shown in Fig. \ref{mhex}. 

Uncertainties have been estimated from the $\chi^{2}$-associated probability as a function of [M/H], after marginalizing the dependence on temperature.
The estimated errors on [M/H] were found in the range $0.06<\sigma_{[M/H]}<0.12~dex$. 
The weighted average of the metallicity turns out to be $\langle[M/H]\rangle=-0.31\pm 0.01$.
The r.m.s. of the sample measures is 0.09 $dex$, similar to the typical individual uncertainties, thus indicating no significant intrinsic spread in metallicity.
The only spectroscopic estimate based on individual spectra is the one by \citet{song2021} who found $\lbrack Fe/H \rbrack = -0.31\pm0.20$, in excellent agreement with the value found here.
A star-to-star comparison among the 9 stars in common indicates $[M/H]_{this~work}-[Fe/H]_{S21}=-0.15\pm0.09~(r.m.s.=0.27)$, which indicates a fair consistency between these two works. 
Other measures available in the literature for this cluster come from integrated spectra of the CaII triplet region 
\citep[$\lbrack Ca/H \rbrack =-0.52\pm0.07$;][]{usher2019} and from Stromgren photometric indices 
\citep[$\lbrack Fe/H \rbrack =-0.53\pm0.04$;][]{piatti2019}, both in reasonable agreement with the value estimated here, considering the systematics possibly affecting estimates obtained with very different tracers and techniques.

The celestial coordinates, WFPC2 magnitudes, gravities, temperatures, radial velocities and metallicities of the 38 targets RSGs of NGC1850A analysed here are listed in Table \ref{rsg_tab}.
  

\subsection{Ba abundance}
\label{ba_sec}

The abundance of barium is the easiest to measure among elements with a dominant s-process contribution, because of its strong and isolated lines which are easily measurable also at relatively low resolution. 
Ba is mainly produced through the slow-neutron capture chain occurring in the envelopes of low- and intermediate-mass asympthotic giant stars.
Its abundance is therefore useful to determine the contribution of low-mass stars to the chemical enrichment of the original medium from which the cluster formed.

Barium abundances have been derived through spectral synthesis calculations of the strong Ba~{\sc ii} line at $6141.7 \AA$ using the driver {\rm synth} in the local thermodynamical equilibrium (LTE) code {\rm MOOG} \citep[][ 2019 version]{sneden1973}. 
The linelist include isotopic splitting and hyperfine structure, following the prescriptions by \cite{mcwilliam1998}. 
We adopted solar-scaled isotopic mixture of $^{134}Ba=2.4\%$, $^{135}Ba=6.6\%$,
$^{136}Ba=7.9\%$, $^{137}Ba=11.2\%$, $^{138}Ba=71.9\%$.
The line under scrutiny in the present study is known to be blended with a Fe~{\sc i} line, which we took into account in our spectral synthesis. 
A metallicity of $[M/H]=-0.3$ was assumed for all our sample stars (see Sect. \ref{met_sec}).
Effective temperature and surface gravity were calculated as described in Sect. \ref{met_sec}, whereas we used the relationship by \citet{kirby2009} to calculate the microturbulent velocities.

For each target, a synthetic spectrum degraded to match the resolution of the observed one has been simulated and the corresponding $\chi^2$ was computed.
The Ba abundance providing the minimum $\chi^{2}$ has been chosen.
Random errors affecting our best-fitting procedure have been evaluated by changing the continuum position until the r.m.s. of the fit around the Ba line was 2.5 times larger than that obtained with the best fit abundance. 
This implies uncertainties between 0.08 - 0.10 dex.
On the other hand, the sensitivities related to the adopted stellar parameters have been evaluated by varying one parameter at the time and inspecting the corresponding
change in abundances \citep[][ and references therein]{dorazi2020}. 
By propagating the uncertainties on these parameters through these sensitivities, an additional scatter of 0.10 dex has been added.

The average Ba abundance derived from the 38 RSG stars turns out to be $\langle [Ba/Fe]\rangle=0.40\pm 0.02$. 
The r.m.s. of the measures is 0.08 dex i.e. compatible with the typical uncertainty in the Ba abundance determination, thus implying no intrinsic spread among the analysed stars.
The Ba abundances of the 38 targets RSGs analysed here are listed in Table \ref{rsg_tab}.

\subsection{Oxygen abundance}
\label{o_sec}

   \begin{figure}
   \centering
   \includegraphics[width=\hsize]{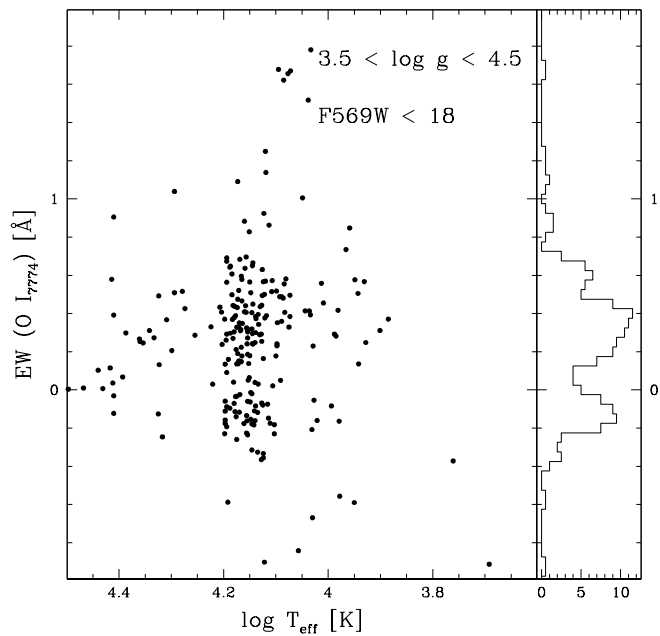}
      \caption{EW of the O~{\sc i} triplet as a function of temperature for bright MS 
      stars. The histograms for stars in the range $4.1<log(T_{eff}/K)<4.2$ in EW are also shown. 
              }
         \label{ot}
   \end{figure}

   \begin{figure}
   \centering
   \includegraphics[width=\hsize]{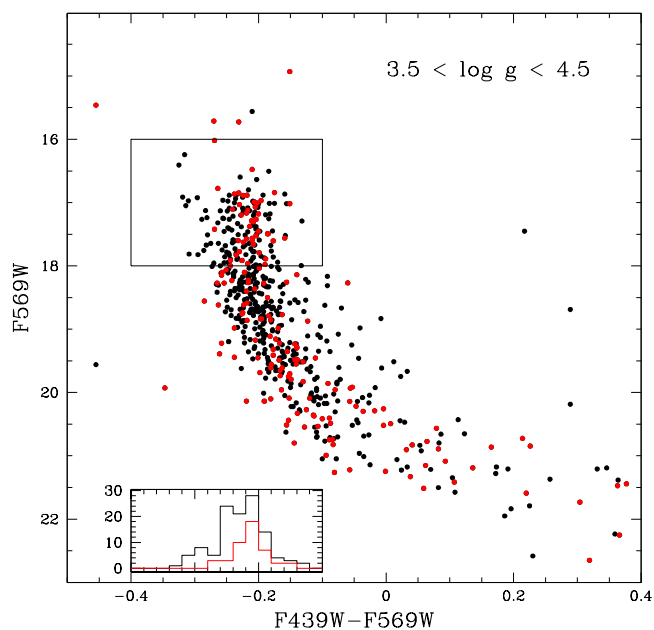}
      \caption{(F439W-F569W)-F569W CMD of NGC1850 stars with $3.5<log~g<4.5$. O-strong and O-weak stars are marked by black and red symbols, respectively.
      The colour histograms of both groups in the magnitude range $16<F569W<18$ are shown in the inset panel.  
              }
         \label{ocm}
   \end{figure}

   \begin{figure}
   \centering
   \includegraphics[width=\hsize]{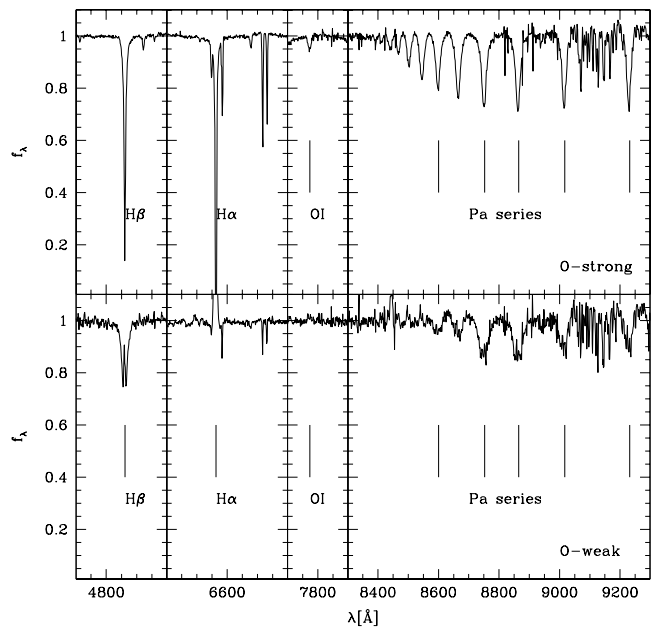}
      \caption{Comparison between the spectra of two O-strong (top panel) and 
      O-weak (bottom panel) MS stars. The H lines as well as the O~{\sc i} triplet are indicated.
              }
         \label{comp}
   \end{figure}
 
   \begin{figure*}
   \centering
   \includegraphics{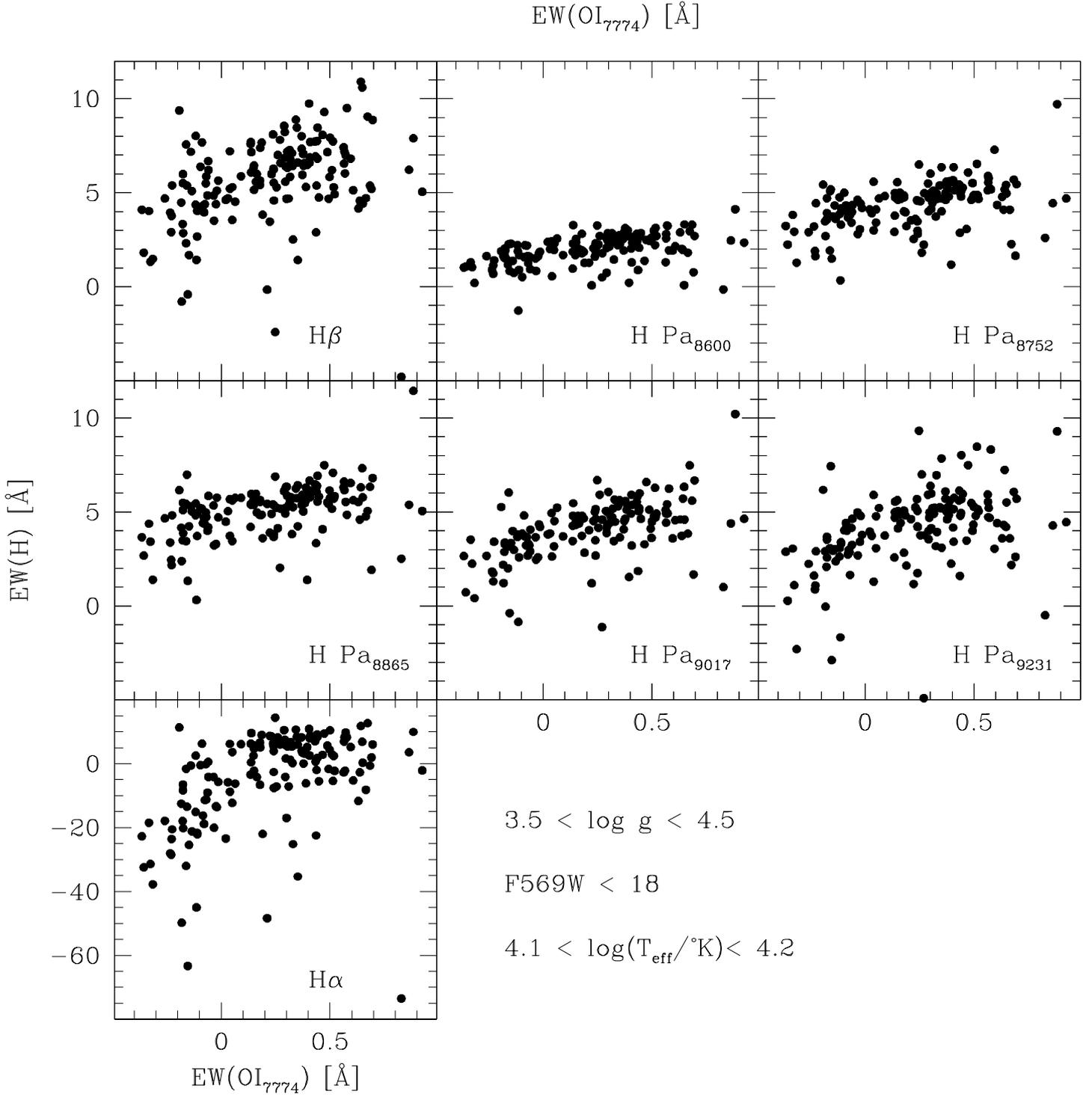}
   \caption{EWs of hydrogen lines as a function of the O~{\sc i} triplet EW for the bright MS stars of NGC1850.}
              \label{oh}%
    \end{figure*}

   \begin{figure*}
   \centering
   \includegraphics[width=\hsize]{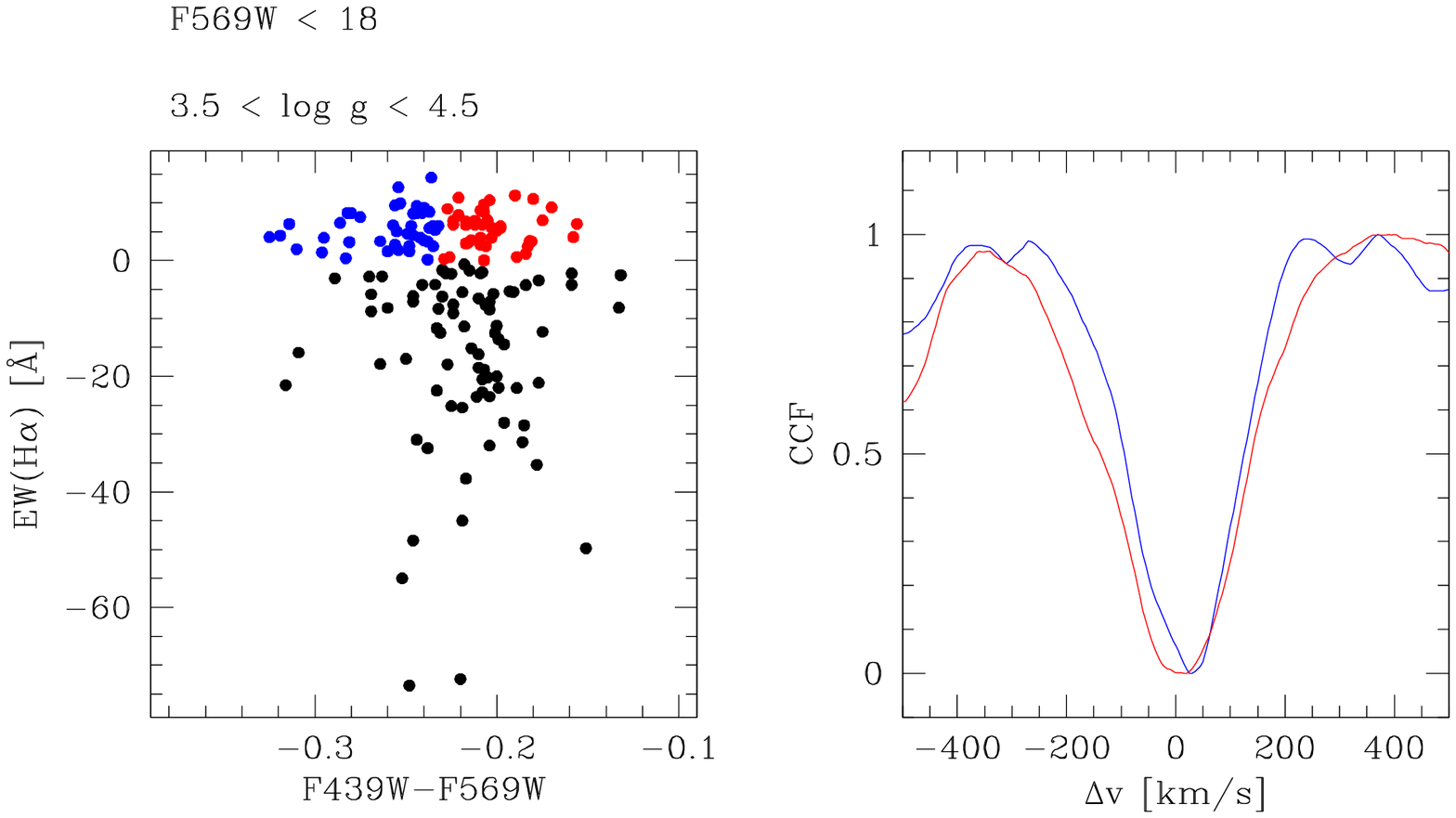}
      \caption{Left panel: Distribution of MSTO stars of NGC1850A with $3.5<log~g<4.5$ and $F569W<18$ in the $F439W-F569W$ vs. $EW(H\alpha)$ plane. 
      Stars bluer and redder than $F439W-F569W=-0.23$ are marked by blue and red points, respectively, while Be stars are marked by black points.
      Right panel: average line profiles of the blue (blue line) and red (red line) samples.  
              }
         \label{hcol}
   \end{figure*}

At odds with RSGs, MS and Blue Supergiants have temperatures $T>8000K$ and contain essentially H and He lines.
However, in the brightest portion of the MS, the S/N is high enough to detect the O~{\sc i} triplet lines at $7771-5\AA$ which appears as a single blended feature at the MUSE resolution. 
This feature has been proven to be extremely useful to determine the O abundance in 
B-F type stars \citep{takeda2008,takeda2016}.

The abundance of oxygen is particularly important for many aspects of stellar physics, cluster formation and chemical enrichment.
Indeed, while metals are produced in both SNe Ia and II, O is mainly released by SNeII and processed through the CNO cycle at high-temperature in massive stars.
The relative fraction of oxygen over iron is therefore large at the beginning of the star formation process and it decreases after the explosion of the first SNe Ia. 
So, the [O/Fe] abundance at a given metallicity can be used as a proxy for the velocity of the chemical enrichment.
Moreover, the presence of an O spread has been the main marker of multiple populations in Galactic globular clusters \citep{carretta2009}.
In rotating massive stars, O spreads can be however produced also by the mixing induced in the outermost layers by meridional circulation.
In these stars, O-depleted material could be dredged-up together with other CNO-processed material \citep{georgy2013}.

We measured the equivalent width of the O~{\sc i} triplet by numerically integrating the 
average continuum-normalized spectrum in the wavelength region between $7765\AA$ and $7785\AA$ bracketing this feature.
The continuum has been set by linearly interpolating the fluxes through two $20\AA$-wide regions on both sides of the O~{\sc i} triplet.
We restrict the analysis of O abundances to those stars along the MS, selecting stars in the gravity range $3.5<log~g<4.5$.
As for any other index, the accuracy in the measure of the O~{\sc i} EW is better at bright magnitudes.

The trend of O~{\sc i} EWs as a function of temperature is shown in Fig. \ref{ot} for the brightest ($F569W<18$) stars. 
As expected, the O~{\sc i} triplet EW shows a clear trend with temperature with a peak at $3.9<log(T_{eff}/K)<4.2$ and a decreasing trend outside this range, reflecting the probability of transition for the neutral oxygen.
However, in the temperature range $4.1<log(T_{eff}/K)<4.2$ a striking bimodality is apparent.
In particular, while the majority of stars in this temperature interval have an $EW\sim0.3\AA$, 
a group of stars constituting $\sim$1/3 of the sample does not host a measurable line. 

A guess of the O abundance from the EW of this line can be determined by interpolating through the non-LTE grid of 
synthetic EWs for this line calculated by \citet{takeda2016} as a function of gravity and temperature.
The group of O-strong stars have a mean abundance $\langle [O/H]\rangle =-0.47\pm 0.04$ (r.m.s.=0.45 dex), corresponding to $\langle [O/Fe]\rangle=-0.16\pm 0.05$.

By applying a conventional separation of the O-strong groups at $EW>0.07\AA$, there are 53 (out of 157; 34\%) O-weak stars brighter than $F569W<18$.
By adopting this classification, we analyse the distribution of these groups in the CMD and across the field of view.
The location in the (F439W-F569W)-F569W CMD is shown in Fig. \ref{ocm}.
It can be noticed that, while at faint magnitudes O-strong and O-weak stars share the same locus on the CMD, close to the MSTO O-weak stars are systematically redder than O-strong ones.
A KS test indicates that the probability that the two colour distribution are drawn from the same parent population is only of 3\%.
We also note that the two groups have similar spatial distributions: a KS test indicates a probability larger that 99\% that the two samples are extracted from the same population.  

A close inspection of individual spectra of O-weak stars reveals that most of them have H lines in emission (see Fig. \ref{comp}).
To quantify this effect, we calculated the EWs of seven H lines.
EWs have been measured on the average spectra, and assuming a $40\AA$ wavelength band around the lines' centers and two contiguous regions of the same width on both sides of the line band to estimate the continuum.
$H\alpha$ in emission ($EW(H\alpha)<0$) has been found in 47 out of 53 (89\%) O-weak stars, while only in 33 out of 104 (32\%) O-strong stars. 
The total fraction of $H\alpha$ emitters ($\sim51\%$) is similar to those measured by \citet{wisniewski2006} and \citet{bastian2017} in the same magnitude range. 
In general, the EWs of all the 7 H lines show a clear increasing trend with the O~{\sc i} triplet EW (see Fig. \ref{oh}). 
This is due to the presence of emission superposed to the canonical line profile which reduces the EW of both H and O~{\sc i} lines \citep{polidan1976}. 
In most cases the emission is often limited to the line core, so that the measured EW, while reduced, remains positive.

\subsection{Rotational velocities}
\label{rot_sec}

As outlined in Sect. \ref{intro_sec}, one of the hypotheses to explain the observed spread in the MSTO of NGC1850 is that it is due to a spread in rotational velocities.

Stellar rotation can be measured in high-resolution spectra through the broadening of isolated spectral lines, measured through the Fourier analysis \citep[see e.g. ][]{diaz2011}.
Unfortunately, the relatively low spectral resolution and S/N of our spectra do not allow to adopt this technique \citep[see ][ for an alternative approach]{kamann2020}.

It is however possible to determine the average line profile of a few isolated lines in the average spectrum of groups of stars selected in colour.
For this purpose the spectra of MSTO stars in the ranges $3.5<log~g<4.5$, $F569W<18$ and $EW(H\alpha)>0~\AA$ (to exclude Be stars, characterized by strongly asymmetric lines) 
have been grouped in two samples according to their colours.
Individual spectra of each group have been corrected for relative velocity and averaged to obtain a high-S/N average spectrum.
The average line profile of a sample of 7 relatively isolated lines (He~{\sc i} 4922, He~{\sc i} 5016, He~{\sc i} 5046/8,  He~{\sc ii} 5411, He~{\sc i} 6678, He~{\sc i} 7065, O~{\sc i} 7771/5),  has been derived by cross correlating the average spectrum with Dirac deltas centered at each line wavelength (se Fig. \ref{hcol}).

It is apparent that the average spectrum of red stars is characterized by broader lines ($FWHM_{red}\sim 280~km~s^{-1}$) than those measured in the average blue one ($FWHM_{blue}\sim 220~km~s^{-1}$), suggesting a higher rotational velocity for red stars.

\section{Discussion}

The analysis of MUSE spectra in the YMC NGC1850 allowed to derive a sound spectroscopic estimate 
of the cluster global metallicity on the basis of individual RSG spectra.
The derived metallicity ($\langle [M/H]\rangle=-0.31\pm0.01$) is more metal rich than the LMC bar \citep[$\lbrack M/H \rbrack=-0.68$, r.m.s.=0.12~dex][]{vanderswaelmen2013} but
similar to that measured in other young LMC clusters \citep{hill1999}.
This is in agreement with the age-metallicity gradient found by many authors \citep{olszewski1991,livanou2013} with young clusters forming from a medium enriched in metals with respect to old stellar populations.
On the other hand, Ba abundances agree with the mean abundance measured in the LMC bar \citep[$\lbrack Ba/Fe \rbrack\sim+0.5$][]{vanderswaelmen2013}, typical of a environments characterized by a slow chemical 
enrichment dominated by the contribution of low-mass Asymptotic Giant Branch stars ($1.5<M/M_{\odot}<4$).

The spectra of RSGs have been also used to determine the systemic velocity and the dynamical mass of NGC1850.
The derived quantities agree with the only available spectroscopic estimates in the 
literature \citep{fischer1993,song2021}. At odds with the study of \citet{fischer1993}, we do not find any significant sign of systemic 
rotation.

We detected a clear bimodality in the EW of the O~{\sc i} triplet at $7771-7775\AA$ among bright MS stars in NGC1850A. 
In particular, the majority of stars display a 
clear absorption while in a group of stars, constituting $\sim 34\%$ of the bright MS sample, this feature is not apparent.
O-weak stars appear slightly redder than O-strong ones, and almost all of them are $H\alpha$ emitters.
The O abundance derived for O-strong stars ($[O/Fe]=-0.16\pm0.05$) agrees with the value measured in stars with similar metallicity in the LMC bar by \citet{vanderswaelmen2013} ($\langle [O/Fe]\rangle\sim-0.2$), characteristic of relatively low-mass galaxes where the overabundance of $\alpha-$elements with respect to Fe produced by SNe has been diluted by the contribution by SNe Ia. 

The most likely interpretation of the observed bimodality is that the EWs of O lines in MSTO stars reflects the degree of rotation of these stars. 
Hydrogen line \citep[and often O~{\sc i} triplet line;][]{polidan1976} emission in hot ($T>10000K$) B stars is indeed often associated with rotation. 
In fact, rapid rotators ($\Omega\sim\Omega_{crit}$) in this phase eject a significant amount of material and develop 
a dense equatorial disk which is excited by the stellar UV photons \citep[Be stars;][]{rivinius2013}. 
The absence of the O~{\sc i} triplet line in O-weak stars could then be 
due to either the superposition of emission and absorption lines \citep{polidan1976} or to a true O depletion in the envelope of these stars driven by rotation-induced mixing \citep{georgy2013}.
The latter hypothesis is however disfavoured by the observational work of \citet{takeda2016} who found no 
dependence of the derived O abundance (calculated using the same indicator we 
adopted in our analysis) from the rotational velocity in the spectra of 34 B stars with different rotational velocities, and by the theoretical models of rotating stars predicting a depletion of only $\Delta[O/H]=-0.03~dex$ for a 4 $M_{\odot}$ star with $\Omega/\Omega_{crit}=0.95$ \citep{georgy2013}.
Following the same line of reasoning, \citet{bastian2017} interpreted the large fraction of Be stars in NGC1850 as an evidence for a large population of rapid rotators.
Rotating stars have an equilibrium structure characterized by larger radii, and consequently lower temperatures and redder colours, 
with respect to non-rotating stars.

The role of rotation in producing the colour spread at the MSTO is supported by the link between line broadening and colour in MSTO stars.
Stars on the red side of the MSTO have indeed broader line profiles, suggesting a higher degree of internal rotation with respect to stars on the blue side \citep[see also][]{kamann2021}.
This could be also the reason why the colour spread remains visible even 
excluding Be stars \citep{correnti2017}: these stars are indeed at the extreme 
of the distribution of rotational velocities ($\Omega\sim\Omega_{crit}$), 
while a shift toward red colours can be present also in stars rotating with 
lower velocities.

An alternative interpretation is that the detected bimodality in O abundance is primordial.
In this hypothesis, it is however not clear the origin of the H lines emission in the oldest O-poor population.

Unfortunately, our spectra have not enough resolution to detect differences in the stellar rotation of 
individual stars.
Future high-resolution spectroscopic analyses focussed on the determination of 
this parameter will be fundamental to clarify the nature of this bimodality.

\begin{acknowledgements}
We warmly thank Alessio Mucciarelli, Livia Origlia and Cristiano Fanelli for useful suggestions.
We also thank Nate Bastian, the referee of our paper, for his helpful comments and suggestions that improved our paper.
Part of this work was supported by the Istituto Nazionale di Astrofisica under the PRIN 2019 founding program 
prop.ID 96 - "Building up the halo: chemo-dynamical tagging in the age of large
surveys" (PI: Lucatello).
\end{acknowledgements}


\begin{thebibliography}{}

\bibitem[Allard(2014)]{allard2014} Allard, F.\ 2014, in "Exploring the Formation and Evolution of Planetary Systems", 299, 271
\bibitem[Asplund et al.(2009)]{asplund2009} Asplund, M., Grevesse, N., Sauval, A.~J. \& Scott, P.\ 2009, \araa, 47, 481
\bibitem[Bacon et al.(2010)]{bacon2010} Bacon, R., Accardo, M., Adjali, L., et al.\ 2010, \procspie, 7735, 773508
\bibitem[Bastian \& de Mink(2009)]{bastian2009} Bastian, N. \& de Mink, S.~E.\ 2009, \mnras, 398, L11
\bibitem[Bastian et al.(2016)]{bastian2016} Bastian, N., Niederhofer, F., Kozhurina-Platais, V., et al.\ 2016, \mnras, 460, L20
\bibitem[Bastian et al.(2017)]{bastian2017} Bastian, N., Cabrera-Ziri, I., Niederhofer, F., et al.\ 2017, \mnras, 465, 4795
\bibitem[Baumgardt \& Hilker(2018)]{baumgardt2018} Baumgardt, H. \& Hilker, M.\ 2018, \mnras, 478, 1520
\bibitem[Brent(2014)]{brent1973} Brent, R.~P.\ 1973, in "Algorithms for minimization without derivatives", Englewood Cliffs, N.J.: Prentice-Hall
\bibitem[Cardelli et al.(1989)]{cardelli1989} Cardelli, J.~A., Clayton, G.~C., \& Mathis, J.~S.\ 1989, \apj, 345, 245
\bibitem[Carretta et al.(2009)]{carretta2009} Carretta, E., Bragaglia, A., Gratton, R.~G., et al.\ 2009, \aap, 505, 117
\bibitem[Correnti et al.(2017)]{correnti2017} Correnti, M., Goudfrooij, P., Bellini, A., Kalirai, J.~S. \& Puzia, T. ~H.\ 2017, \mnras, 467, 3628
\bibitem[Dalessandro et al.(2018)]{dalessandro2018} Dalessandro, E., Zocchi, A., Varri, A.~L., et al.\ 2018, \mnras, 474, 2277
\bibitem[D'Ercole et al.(2008)]{dercole2008} D'Ercole, A., Vesperini, E., D'Antona, F., McMillan, S. ~L.~W. \& Recchi, S.\ 2008, \mnras, 391, 825
\bibitem[D{\'\i}az et al.(2011)]{diaz2011} D{\'\i}az, C.~G., Gonz{\'a}lez, J.~F., Levato, H. \& Grosso, M. \ 2011, \aap, 531, A143
\bibitem[Dolphin(2000)]{dolphin2000} Dolphin, A.~E.\ 2000, \pasp, 112, 1383
\bibitem[D'Orazi et al.(2020)]{dorazi2020} D'Orazi, V., Oliva, E., Bragaglia, A., et al.\ 2020, \aap, 633, A38
\bibitem[Fischer et al.(1993)]{fischer1993} Fischer, P., Welch, D.~L., \& Mateo, M.\ 1993, \aj, 105, 938
\bibitem[Georgy et al.(2013)]{georgy2013} Georgy, C., Ekstr{\"o}m, S., Granada, A., et al.\ 2013, \aap, 553, A24
\bibitem[Gilmozzi et al.(1994)]{gilmozzi1994} Gilmozzi, R., Kinney, E.~K., Ewald, S.~P., Panagia, N. \& Romaniello, M.\ 1994, \apjl, 435, L43
\bibitem[G{\'o}rski et al.(2020)]{gorski2020} G{\'o}rski, M., Zgirski, B., Pietrzy{\'n}ski, G., et al.\ 2020, \apj, 889, 179
\bibitem[Hill \& Spite(1999)]{hill1999} Hill, V. \& Spite, M.\ 1999, \apss, 265, 469
\bibitem[Kamann et al.(2020)]{kamann2020} Kamann, S., Bastian, N., Gossage, S., et al.\ 2020, \mnras, 492, 2177
\bibitem[Kamann et al.(2021)]{kamann2021} Kamann, S., Bastian, N., Usher, C., Cabrera-Ziri, I. \& Saracino, S.\ 2021, \mnras
\bibitem[King(1966)]{king1966} King, I.~R.\ 1966, \aj, 71, 64
\bibitem[Kirby et al.(2009)]{kirby2009} Kirby, E.~N., Guhathakurta, P., Bolte, M., Sneden, C. \& Geha, M.~C.\ 2009, \apj, 705, 328
\bibitem[Livanou et al.(2013)]{livanou2013} Livanou, E., Dapergolas, A., Kontizas, M., et al.\ 2013, \aap, 554, A16
\bibitem[Mackey et al.(2008)]{mackey2008} Mackey, A.~D., Broby Nielsen, P., Ferguson, A.~M.~N. \& Richardson, J.~C.\ 2008, \apjl, 681, L17
\bibitem[Marigo et al.(2008)]{marigo2008} Marigo, P., Girardi, L., Bressan, A., et al.\ 2008, \aap, 482, 883
\bibitem[McLaughlin \& van der Marel(2005)]{mclaughlin2005} McLaughlin, D.~E. \& van der Marel, R.~P.\ 2005, \apjs, 161, 304
\bibitem[McWilliam(1998)]{mcwilliam1998} McWilliam, A.\ 1998, \aj, 115, 1640
\bibitem[Milone et al.(2009)]{milone2009} Milone, A.~P., Bedin, L.~R., Piotto, G. \& Anderson, J.\ 2009, \aap, 497, 755
\bibitem[Mucciarelli et al.(2012)]{mucciarelli2012} Mucciarelli, A., Origlia, L., Ferraro, F.~R., Bellazzini, M. \& Lanzoni, B.\ 2012, \apjl, 746, L19
\bibitem[Niederhofer et al.(2015)]{niederhofer2015} Niederhofer, F., Hilker, M., Bastian, N. \& Silva-Villa, E.\ 2015, \aap, 575, A62
\bibitem[Olszewski et al.(1991)]{olszewski1991} Olszewski, E.~W., Schommer, R.~A., Suntzeff, N.~B. \& Harris, H.~C.\ 1991, \aj, 101, 515
\bibitem[Piatti et al.(2019)]{piatti2019} Piatti, A.~E., Pietrzy{\'n}ski, G., Narloch, W., G{\'o}rski, M. \& Graczyk, D.\ 2019, \mnras, 483, 4766
\bibitem[Piotto et al.(2015)]{piotto2015} Piotto, G., Milone, A.~P., Bedin, L.~R., et al.\ 2015, \aj, 149, 91
\bibitem[Polidan \& Peters(1976)]{polidan1976} Polidan, R.~S. \& Peters, G.~J.\ 1976, in "Be and Shell Stars", IAUS, 70, 59
\bibitem[Portegies Zwart et al.(2010)]{portegieszwart2010} Portegies Zwart, S.~F., McMillan, S.~L.~W., \& Gieles, M.\ 2010, \araa, 48, 431
\bibitem[Pryor \& Meylan(1993)]{pryor1993} Pryor, C. \& Meylan, G.\ 1993, Structure and Dynamics of Globular Clusters, 50, 357
\bibitem[Rivinius et al.(2013)]{rivinius2013} Rivinius, T., Carciofi, A.~C., \& Martayan, C.\ 2013, \aapr, 21, 69
\bibitem[Sneden(1973)]{sneden1973} Sneden, C.~A.\ 1973, Ph.D. Thesis
\bibitem[Song et al.(2021)]{song2021} Song, Y.-Y., Mateo, M., Bailey, J.~I., et al.\ 2021, \mnras, 504, 4160
\bibitem[Takeda \& Honda(2016)]{takeda2016} Takeda, Y. \& Honda, S.\ 2016, \pasj, 68, 32
\bibitem[Takeda et al.(2008)]{takeda2008} Takeda, Y., Han, I., Kang, D.-I., Lee, B.-C. \& Kim, K.-M.\ 2008, Journal of Korean Astronomical Society, 41, 83
\bibitem[Tody(1986)]{tody1986} Tody, D.\ 1986, \procspie, 627, 733
\bibitem[Usher et al.(2019)]{usher2019} Usher, C., Beckwith, T., Bellstedt, S., et al.\ 2019, \mnras, 482, 1275
\bibitem[Van der Swaelmen et al.(2013)]{vanderswaelmen2013} Van der Swaelmen, M., Hill, V., Primas, F. \& Cole, A.~A. \ 2013, \aap, 560, A44
\bibitem[Weilbacher et al.(2020)]{weilbacher2020} Weilbacher, P.~M., Palsa, R., Streicher, O., et al.\ 2020, \aap, 641, A28
\bibitem[Wisniewski \& Bjorkman(2006)]{wisniewski2006} Wisniewski, J.~P. \& Bjorkman, K.~S.\ 2006, \apj, 652, 458
\end{thebibliography}
\end{document}